\documentstyle[epsfig,twoside,fleqn,espcrc2,axodraw]{article}

\def\etal{\hbox{\it et al., }}
\def\lsim{\raise0.3ex\hbox{$\;<$\kern-0.75em\raise-1.1ex\hbox{$\sim\;$}}}
\def\gsim{\raise0.3ex\hbox{$\;>$\kern-0.75em\raise-1.1ex\hbox{$\sim\;$}}}

\newcommand{\AmS}{{\protect\the\textfont2
  A\kern-.1667em\lower.5ex\hbox{M}\kern-.125emS}}

\title{
\vskip-1.3cm\rightline{\small{hep-ph/9911274}}
\vskip-0.2cm\rightline{\small{UCCHEP/5-99}}
\vskip-0.2cm\rightline{\small{FSU-HEP-991020}}
\vskip+0.1cm
Stop Decays with R--Parity Violation and the Neutrino Mass}

\author{Marco Aurelio D\'\i az
                              \address{Facultad de F\'\i sica,
        Universidad Cat\'olica de Chile, 
        Av. Vicu\~na Mackenna 4860, Santiago, Chile, and \\
\hskip0.2cm Department of Physics,
        Florida State University, 
        Tallahassee, Florida 32306, USA
\\ \vskip0.4cm Talk given at the International Workshop on Particles in 
            Astrophysics and Cosmology: From Theory to Observation,
            Valencia, Spain, May 3-8, 1999.
}}
       
\begin{document}

\begin{abstract}
The atmospheric and solar neutrino problems can be explained in a 
supersymmetric  scenario where R--parity is broken bilinearly. Within this
context we explore the decays of the top squark. We find that the Rp violating
decay $\tilde t_1\rightarrow b\tau$ can easily dominate over the Rp
conserving decay $\tilde t_1\rightarrow c\tilde\chi^0_1$ and sometimes also 
over the decay $\tilde t_1\rightarrow b\tilde\chi^+_1$. We study the 
implications of non--universal boundary conditions at the GUT scale.
\end{abstract}

\maketitle

The SuperKamiokande collaboration has confirmed the deficit of
muon neutrinos from atmospheric neutrino data \cite{SuperK}. The simplest 
interpretation of the data is through $\nu_{\mu}$ to $\nu_{\tau}$
flavour oscillations with maximal mixing. This experimental observation
has profound implications on one of the fundamental problems in modern
physics, namely, the pattern of fermion masses and mixings, and the origin
of mass. Recently, a supersymmetric solution to this problem was
proposed \cite{RDHPV}, based on an extension of the Minimal Supersymmetric 
Standard Model (MSSM) where R--Parity and lepton number are violated 
through bilinear terms in the superpotential (BRpV) \cite{brpv,brpvus}. 
The model can be embedded into supergravity (SUGRA) with universal boundary 
conditions at the Grand Unification scale (GUT), and radiative electroweak 
breaking \cite{DRV}, in which case, this is a one parameter extension of 
Minimal SUGRA. 

The superpotential contains the following bilinear terms
\begin{equation}
W=-\mu\widehat H_d\widehat H_u+\epsilon_i\widehat L_i\widehat H_u+...
\label{BSup}
\end{equation}
where $\epsilon_i$ are BRpV parameters with units of mass. The rest of the
superpotential corresponds to the usual Rp conserving Yukawa terms. A tree
level neutrino mass is generated through neutrino mixing with neutralinos 
and a see-saw type mechanism. If the $\epsilon_i$ are small, the tree level
neutrino mass can be approximated to 
\begin{equation}
m_{\nu}\approx{{g^2M_1+g'^2M_2}\over{4{\mathrm{det}}({\bf M}_{\chi^0})}}
|\vec\Lambda|^2
\label{numass}
\end{equation}
where $\Lambda_i=\mu v_i+v_d\epsilon_i$, and $v_i$ are the sneutrino vevs
\cite{HV,RDHPV}. The parameters $\Lambda_i$ are proportional to the 
vevs of the sneutrinos in the basis where $\epsilon$ terms are removed from 
the superpotential. The $\Lambda_i$ play a crucial role in the determination 
of the atmospheric angle. For example, 
\begin{figure}[htb]
\vspace{-6cm}
\centerline{ \hskip 2.7cm \epsfxsize 4.0 truein \epsfbox {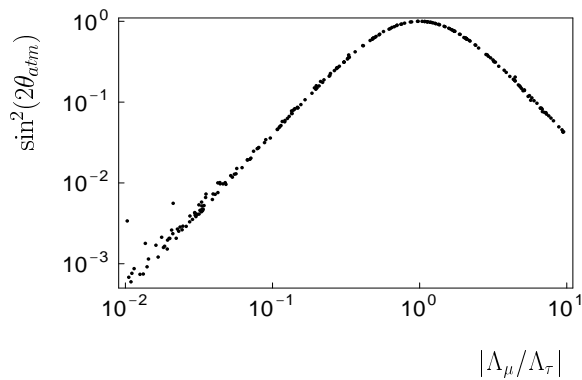}} 
\vskip 0.2cm
\caption{Atmospheric angle as a function of $|\Lambda_{\mu}/\Lambda_{\tau}|$
for $\Lambda_e=0.1\Lambda_{\tau}$.}
\vskip -0.5cm
\label{lettfig1}
\end{figure}
in Fig.~\ref{lettfig1} we have the atmospheric angle as a function of 
$|\Lambda_{\mu}/\Lambda_{\tau}|$ for $|\Lambda_e|=0.1|\Lambda_{\tau}|$. 
Maximality is obtained for $|\Lambda_{\mu}|\approx|\Lambda_{\tau}|$ as long
as $|\Lambda_e|$ is small. The solar neutrino problem can also be 
solved by this mechanism \cite{RDHPV}.

In the determination of neutrino masses and mixing angles the one--loop 
contributions are essential \cite{ralf,RDHPV}. The main loops are given by

\begin{picture}(180,40)(0,20) 
\ArrowLine(0,25)(25,25)
\ArrowArcn(40,25)(15,180,0)
\DashCArc(40,25)(15,180,0){3}
\ArrowLine(55,25)(80,25)
\Text(90,25)[]{$+$}
\ArrowLine(100,25)(125,25)
\ArrowArcn(140,25)(15,180,0)
\PhotonArc(140,25)(15,180,0){2}{8.5}
\ArrowLine(155,25)(180,25)
\Text(12,32)[]{$\nu$}
\Text(67,32)[]{$\nu$}
\Text(112,32)[]{$\nu$}
\Text(167,32)[]{$\nu$}
\Text(45,48)[]{$b,\tilde\chi^+$}
\Text(45,3)[]{$\tilde b, H^-$}
\Text(145,48)[]{$\tilde\chi^+$}
\Text(145,3)[]{$W^-$}
\end{picture}
\vskip 0.8cm
\noindent
where the bottom--sbottom contribution is the most important one. One--loop 
contributions increase as $\tan\beta$ increases and as $m_0$ decreases
\cite{RDHPV}. Charginos mix with charged sleptons, therefore the last ones 
also contribute, although less importantly. The phenomenological implications 
of this mixing in the tau sector can be found in \cite{ADV}. In particular, 
the mixing does not spoil the well measured $Z\tau\tau$ and $W\tau\nu_{\tau}$ 
couplings.

In BRpV neutral Higgs bosons mix with sneutrinos. This mixing does not 
affect the upper bound on the neutral CP-even Higgs mass, although in 
general lowers $m_h$ in a few GeV \cite{DRV}. Interestingly, the SUSY 
solution to the atmospheric and solar neutrino problems requires low 
$\tan\beta$ which in turn implies low Higgs mass $m_h$ \cite{DiazHaberii}. 
In fact, it was found that $\tan\beta\lsim10$ and $m_h\lsim115$ GeV 
\cite{RDHPV} is required to solve the neutrino problems. Small values of 
$\tan\beta$ are already been explored by neutral Higgs searches, 
preliminary ruling out $1<\tan\beta\lsim1.8$ \cite{alephmh}.

\begin{figure}[htb]
\vspace{-0.7cm}
\centerline{ \hskip .6cm \epsfxsize 2.5 truein \epsfbox {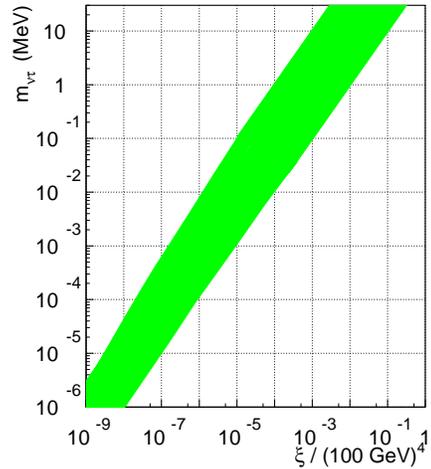}} 
\vskip -0.7cm
\caption{Tau--neutrino mass at tree level as a function of the parameter
$\xi=\Lambda_{\tau}^2$.}
\vskip -0.4cm
\label{mnt_xi_new}
\end{figure}
Beside the effects on the neutrino sector, BRpV has important implications
on the phenomenology of supersymmetric particles. To study them, it is 
usually a good approximation to consider BRpV only in the tau sector:
$\epsilon_1=\epsilon_2=0$, $\epsilon_3\ne0$. In this case, the tau-neutrino
mass is the only non--zero, is given by 
\begin{equation}
m_{\nu_{\tau}}\approx{{g^2}\over{2M}}v'^2_3
\label{taunumass}
\end{equation}
and depends only on the tau--sneutrino vev $v'_3$ in the rotated basis. In 
this way, $m_{\nu_{\tau}}$ gives a good order of magnitude of the neutrino 
masses in the complete theory when the details in the neutrino sector are not 
relevant. We have made a scan over all the parameters of the theory and in 
Fig.~\ref{mnt_xi_new} we plot the tau--neutrino mass as a function of the
parameter $\xi\equiv\Lambda_{\tau}^2$. We appreciate that $m_{\nu_{\tau}}$
values can be obtained from the collider upper bound of 18 MeV 
\cite{alephnutau} down to eV or smaller.

The phenomenology of the top--squarks (stops) is also affected by BRpV
\cite{stop,morestop}. The Rp violating couplings of the stop are governed 
by $\epsilon_3$ and not by the neutrino mass, which makes them potentially 
large. In the following we study in detail the Rp violating decay mode 
$\tilde t_1\rightarrow b\tau$ and compare it with the Rp conserving decays 
$\tilde t_1\rightarrow c\chi^0_1$ and $\tilde t_1\rightarrow b\chi^+_1$.
\begin{figure}[htb]
\vspace{-0.7cm}
\centerline{ \hskip .6cm \epsfxsize 3.5 truein \epsfbox {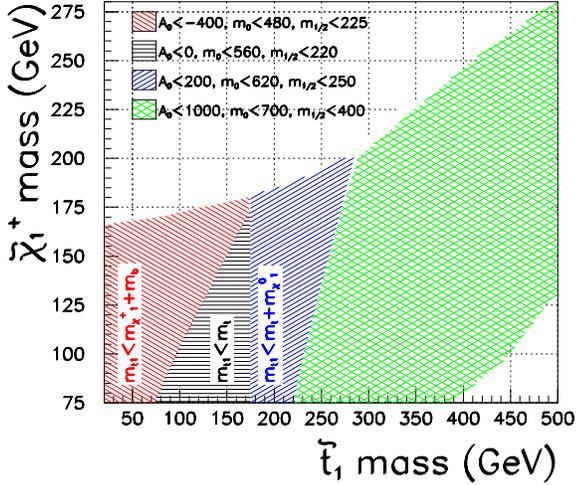}} 
\vskip -0.7cm
\caption{Kinematical regions in the stop--chargino mass plane relevant for
the decay modes $\tilde t_1\rightarrow b\tau$, 
$\tilde t_1\rightarrow c\chi^0_1$, and $\tilde t_1\rightarrow b\chi^+_1$.}
\vskip -0.4cm
\label{paramspace}
\end{figure}
In Fig.~\ref{paramspace} we plot the regions of parameter space relevant to 
the three decay modes mentioned. We work with BRpV embedded into SUGRA with
universal scalar $m_0$ and gaugino $M_{1/2}$ masses at the GUT scale.

We do not study here the decay mode $t\rightarrow bH^+$. The charged Higgs 
mixes with the staus and their rich phenomenology in BRpV can be found in
\cite{ADFGV}. In particular, a charged Higgs lighter than in the MSSM, even
after radiative corrections \cite{diazhaberi}, can be found in global BRpV. 
In BRpV--SUGRA though, the charged Higgs is usually heavier. In addition, 
the constraints on the charged Higgs mass from the CLEO measurement for
$B(b\rightarrow s\gamma)$ \cite{CLEObsg} in the MSSM \cite{bsgMSSM} are
relaxed in BRpV \cite{bsgBRpV}.

The first issue we point out is that the one step approximation of the RGE's
in the analytic determination of 
$\Gamma(\tilde t_1\rightarrow c\tilde\chi^0_1)$ \cite{HK} is usually off 
by one order of magnitude or more.
\begin{figure}[htb]
\vspace{-0.7cm}
\centerline{ \hskip .6cm \epsfxsize 3.5 truein \epsfbox {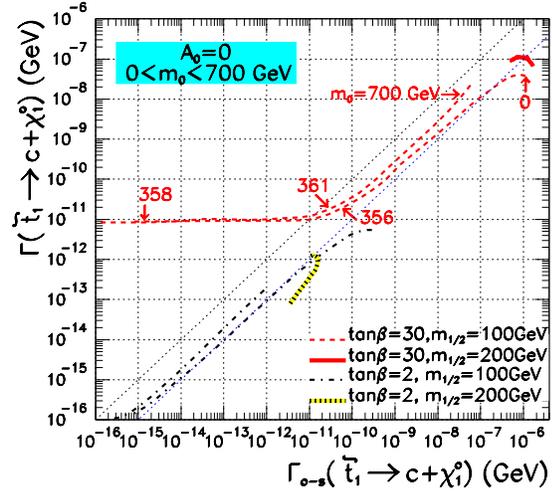}} 
\vskip -0.7cm
\caption{Comparison between the one step approximation of 
$\Gamma(\tilde t_1\rightarrow c\tilde\chi^0_1)$ and its exact numerical
value for different regions in parameter space.}
\vskip -0.4cm
\label{gtn2c-gtn2cos}
\end{figure}
In Fig.~\ref{gtn2c-gtn2cos} we compare the exact solution \cite{stop} with
the one--step approximation for four different values of $\tan\beta$ and
$M_{1/2}$. Note that the approximation can fail for several orders of
magnitude for one of the examples. The main reason for this behaviour is
that the evolution of $A_b$ depends strongly on $M_{1/2}$ and $\tan\beta$,
and this dependence is lost in the one--step approximation.

\begin{figure}[htb]
\vspace{-0.7cm}
\centerline{ \hskip .6cm \epsfxsize 3.5 truein \epsfbox {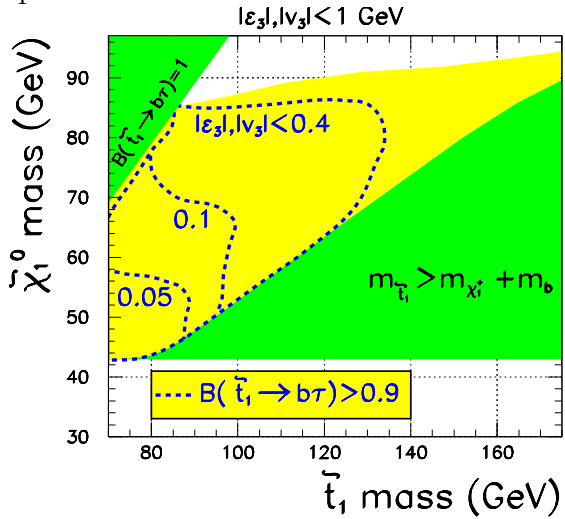}} 
\vskip -0.7cm
\caption{Regions in the $m_{\tilde\chi^0_1}-m_{\tilde t_1}$ plane where
the Rp violating decay mode $\tilde t_1\rightarrow b\tau$ is larger than
90\%.}
\vskip -0.4cm
\label{x1-t1-1}
\end{figure}
In Fig.~\ref{x1-t1-1} we compare the branching ratios of the Rp violating
decay mode $\tilde t_1\rightarrow b\tau$ and the Rp conserving decay
$\tilde t_1\rightarrow c\chi^0_1$ in the regions of parameter space where
the decay $\tilde t_1\rightarrow b\chi^+_1$ is closed. We observe that 
even for small values of the BRpV parameters $\epsilon_3$ and $v_3$ it is
very easy for the Rp violating decay mode to dominate: 
$B(\tilde t_1\rightarrow b\tau)>0.9$. This has crucial consequences in the
experimental strategies in the search for top squarks in the region where
the stop is roughly lighter than the chargino. We note that
$\tilde t_1\rightarrow b\tau$ can dominate even for neutrino masses of
the order of $10^{-2}<m_{\nu_{\tau}}<10^{-1}$ eV \cite{stop}.

\begin{figure}[htb]
\vspace{-0.7cm}
\centerline{ \hskip .6cm \epsfxsize 3.5 truein \epsfbox {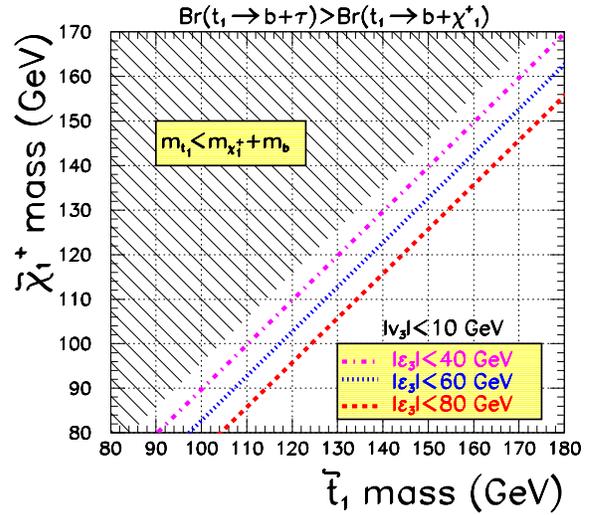}} 
\vskip -0.7cm
\caption{Contours where the decay $\tilde t_1 \rightarrow b\tau$ dominates
over $\tilde t_1 \rightarrow b\tilde\chi_1^+$ in the plane $m_{\tilde t_1}$
v/s $m_{\tilde\chi_1^+}$.}
\vskip -0.4cm
\label{x1-t1-z2}
\end{figure}
If $m_{\tilde t_1}>m_{\tilde\chi^+_1}+m_b$ then the stop can decay also into
a chargino and a bottom quark. In Fig.~\ref{x1-t1-z2} we compare this decay
mode with the Rp violating one. We show in the plane 
$m_{\tilde t_1}-m_{\tilde\chi_1^+}$ the regions where the RpV decay 
$B(\tilde t_1 \rightarrow b\tau)$ dominates over the Rp conserving decay 
$B(\tilde t_1 \rightarrow b\tilde\chi_1^+)$. In the shaded region the 
decay $\tilde t_1 \rightarrow b\tilde\chi_1^+$ is not allowed. Between 
the shaded region and the inclined lines we have an RpV decay rate
larger than the Rp conserving one. For small values of the parameter
$\epsilon_3$ the RpV decay dominates only if close to the threshold, where
a large kinematical suppression is present.

It is interesting to correlate these branching ratios with the neutrino mass,
which in first approximation it is given by eq.~(\ref{taunumass}). The 
sneutrino vev is determined by the minimization of the scalar potential,
which in BRpV involve three tadpole equations. In models with universality of 
scalar masses at the GUT scale the sneutrino vev at tree level is \cite{talks}
\begin{equation}
v'_3\approx -{{\epsilon_3\mu}\over{\mu'^2m_{\tilde\nu^0_{\tau}}^2}}
\left(v'_1\Delta m^2+\mu'v_2\Delta B\right)
\label{sneuvev}
\end{equation}
where $\Delta m^2=m_{H_1}^2-m_{L_3}^2$ and $\Delta B=B_3-B$ are evaluated at 
the weak scale. With soft universality at the GUT scale, these two differences
are naturally small because they are radiativelly generated and proportional
to the bottom Yukawa coupling squared. In order to have small neutrino masses
without small values of $\epsilon_3$ it is some times necessary to rely in 
cancelations between the two terms in eq.~(\ref{sneuvev}). In the next 
figures we see that this cancelation is not big enough to consider it a fine
tunning. In addition, we study the effect of relaxing the universality at 
the GUT scale.

\begin{figure}[htb]
\vspace{-0.7cm}
\centerline{ \hskip .6cm \epsfxsize 3.5 truein \epsfbox {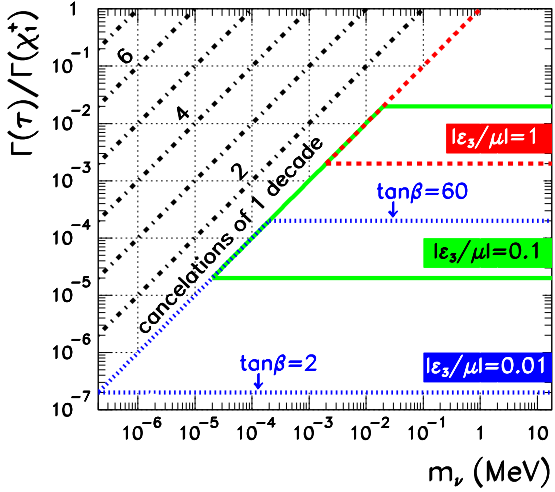}} 
\vskip -0.7cm
\caption{Allowed regions for $\Gamma(\tilde t_1 \rightarrow b\tau)/
\Gamma(\tilde t_1 \rightarrow b\tilde\chi_1^+)$ as a function of the tau 
neutrino mass for different amounts of cancellation between the two terms
that contribute to $m_{\nu_{\tau}}$. We consider $|\epsilon_3/\mu|=1$
(inside the dashed lines), $|\epsilon_3/\mu|=0.1$ (solid), and 
$|\epsilon_3/\mu|=0.01$ (dotted).}
\vskip -0.4cm
\label{gtau-gchar-nu}
\end{figure}
In Fig.~\ref{gtau-gchar-nu} we have the ratio $\Gamma(\tilde t_1 
\rightarrow b\tau)/\Gamma(\tilde t_1 \rightarrow b\tilde\chi_1^+)$ as a 
function of the neutrino mass. Large Rp violating branching ratios appear
for large $\tan\beta$ and large values of $\epsilon_3$. Larger 
$\Gamma(\tilde t_1  \rightarrow b\tau)$ and smaller neutrino masses are 
obtained if we accept cancellations between the two terms in 
eq.~(\ref{sneuvev}). We do not think that cancelations up to four
order of magnitude is a fine tunning, as it happens between vev's in the 
MSSM with large values of $\tan\beta$. We have omitted from 
Fig.~\ref{gtau-gchar-nu} points with large kinematical suppressions.

In Fig.~\ref{gtau-gchar-nu} we imposed $m_{H_1}^2=m_{L_3}^2$ at the GUT scale
but not $B_3=B$. The effect of imposing universality $B_3=B$ at the GUT scale
is to eliminate the points in the lower right corner in 
Fig.~\ref{gtau-gchar-nu} \cite{stop}. The effect of not imposing the 
universality condition $m_{H_1}^2=m_{L_3}^2$ at the GUT scale is more 
dramatic and is analyzed in the next figure.

\begin{figure}[htb]
\vspace{-0.7cm}
\centerline{ \hskip .6cm \epsfxsize 3.5 truein \epsfbox {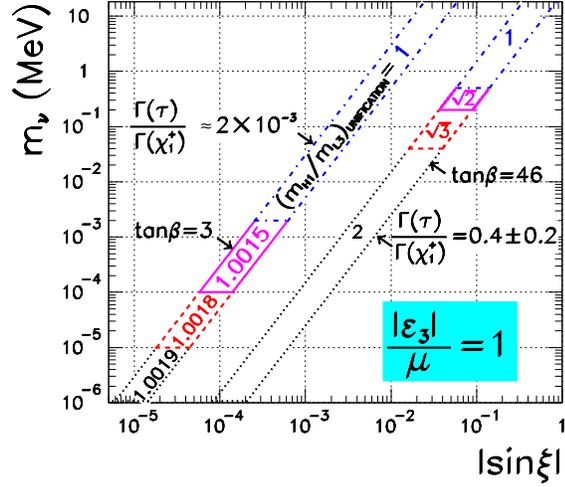}} 
\vskip -0.7cm
\caption{Neutrino mass as a function of $\sin\xi=v'_3/v'_1$ and the effect 
of non--universality $m_{H_1}^2/m_{L_3}^2\ne1$ at the GUT scale.}
\vskip -0.4cm
\label{mnu-sinxi}
\end{figure}
In Fig.~\ref{mnu-sinxi} we plot the neutrino mass as a function of the 
parameter $\sin\xi=v'_3/v'_1$ for $|\epsilon_3|/\mu=1$ and the two values 
$\tan\beta=3$ and 46, which translates into the nearly constant values
$\Gamma(\tilde t_1 \rightarrow b\tau)/\Gamma(\tilde t_1 \rightarrow 
b\tilde\chi_1^+)=2\times10^{-3}$ and $0.4\pm0.2$ respectively. We are 
accepting cancelations between the two terms in eq.~(\ref{sneuvev}) of only 
one order of magnitude. The minimum value of $m_{\nu_{\tau}}$ attainable in 
each case depends on the degree of non--universality. For $\tan\beta=3$ tiny 
deviations from $m_{H_1}^2/m_{L_3}^2=1$ are enough to obtain neutrino masses
of the order of eV. For $\tan\beta=46$ larger deviations are necessary.
We note that SUGRA models based on the $SO(10)$ gauge group with universality
of Yukawa couplings, which need values of $\tan\beta\sim45-55$, can produce 
the correct amount of non--universality from extra D-term contributions
associated with the reduction in rank of the gauge symmetry group when 
spontaneously breaks to $SU(3)\times SU(2)\times U(1)$ \cite{so10}.

In summary, the Bilinear R--Parity Violating extension of the MSSM can solve
the atmospheric and solar neutrino problems by giving mass and mixing to the
neutrinos through mixing with neutralinos. In this context, the Rp violating
decay mode of the lightest stop $\tilde t_1\rightarrow b\tau$ can dominate
over the conventional modes introducing new signals that modofy the search
of stops in colliders.

\section*{Acknowledgments}

I am thankful to  my collaborators A. Akeroyd, J. Ferrandis, 
M.A. Garcia-Jare\~no, M. Hirsch, W. Porod, J.C. Rom\~ao, E. Torrente-Lujan, 
J.W.F. Valle, and specially D.A. Restrepo, without whom this work would not 
have been possible.

\end{document}